# Polarization and entanglement in spin systems

*G.B. Furman[1,2], V. M. Meerovich[1], and V.L. Sokolovsky[1]*

[1]Department of Physics, Ben-Gurion University of the Negev, POB 653, Beer-Sheva 84105, Israel

[2]Ohalo College, POB 222, Qazrin 12900, Israel

October 23, 2008

We established important relationships between entanglement measures and the order parameter (spin polarization) in nuclear spin systems controlled by the nuclear magnetic resonance (NMR) technique. Since spin polarization can be easily manipulated by the NMR technique, experimentalists are presented with an opportunity to study the dynamic properties of entanglement, i.e., the creation and evolution of entangled states. Our approach may constitute the basis for researching the relations between the entanglement measures and measurable parameters of order in other quantum systems.

The important role of quantum entanglement [1-4] as a resource in quantum computing [5], quantum communication [6], and quantum metrology [7, 8] has stimulated intensive qualitative and quantitative research. The static properties of the entanglement in spin chains has been extensively studied [3-4]. However, to understand the creation and transport of entanglement of the system with the number of qubits, it is of great interest to study the entanglement dynamics between pairs of qubits. Recently, the study was extended to the understanding of entanglement dynamics in some model systems [3-4]. One of the important tasks is determination of measures of entanglement [2-4]. Several criteria have been proposed to distinguish separable from entangled states and to quantify entanglement. Various entanglement witnesses have recently been obtained in terms of expectation values of macroscopic observables such as internal energy [9, 10], magnetization [11], and magnetic susceptibility [12]. However, those results were obtained for systems at thermal equilibrium at very low temperatures and cannot be used for investigations of entanglement dynamics. The most natural measures of entanglement are the von Neumann entropy [13, 14], which was introduced for pure states and in a static regime, and the concurrence, which is used to calculate the entanglement of the formation of two-qubit systems. However, it is difficult to directly measure both entropy and concurrence [15], especially under dynamic conditions.

The entropy in thermodynamics can be related to the order parameter [16-18], which in many cases is a directly measurable value. In spin systems under the order parameter can be used the polarization $P$, for ferromagnetic or the polarization of the sublattice for antiferromagnetic [16]. Expanding the entropy to second order in the order parameter at a vicinity of a point of phase transition from a phase *I* to a phase *II* the following expression for the entropy can be obtained [16]

$$S = S_0 - \frac{\partial A}{\partial T} P^2, \qquad (1)$$

where $S_0$ is the entropy in phase $I$ and in the $A$ is function of the temperature $T$. The entropy for non-interacting spins ½ with the polarization takes the form [17, 18]:

$$S = 2\ln 2 - (1+P)\ln(1+P) - (1-P)\ln(1-P). \qquad (2)$$

For pure quantum state in static regime, the concurrence of a two-qubit can be written in terms of the determinant of the reduce density matrix [19] and the concurrency can be connected with polarization of one spin [3]. The von Neumann entropy is also relevant to the energy and other conserved quantities that appear in macroscopic thermodynamics [20]. The question that we pose therefore is: Can the von Neumann entropy, which describes the degree of entanglement in a quantum system, also be related to a measurable order parameter for time dependent state? To answer this question, we consider a linear chain of $N$ nuclear spins coupled by long-range dipolar interactions and subjected to an external magnetic field, $\vec{H}_0 = H_0\vec{z}$. We use selective transverse irradiation along the $x$-axis [21, 22] and assume that only two spins, $m$ and $n$, are simultaneously irradiated at their resonance frequencies. The Hamiltonian of the model is:

$$H(t) = \sum_{k=1}^{N}\left\{\omega_{0k}I_k^z + 2(\delta_{k,n} + \delta_{k,m})\omega_{1k}I_k^x \cos\omega_0 t\right\} + H_{dd} \qquad (3)$$

where $\omega_{0k} = \gamma_k H_0$ is the energy difference between the excited and ground states of an isolated spin, $\gamma_k$ is the gyromagnetic ratio of the $k$-th spin, $\omega_{1k}$ is the amplitude of irradiation field acting on the $k$-th spin, $I_k^x$ and $I_k^z$ are the projections of the angular spin momentum operators $\vec{I}_k$ on the $x$- and $z$- axes, respectively, and $\delta_{m,n}$ is the Kronecker delta. The Hamiltonian $H_{dd}$ is describing dipolar interactions. In the rotating frame the fast oscillating terms with frequencies $\omega_{0k}$ and $2\omega_{0k}$ can be omitted [23, 24], and the Hamiltonian (1) reduces to the form:

$$H = \sum_{k=1}^{N} \left\{ (\delta_{k,n} + \delta_{k,m}) \omega_{1k} I_k^x \right\} + H_{dd}^{sec}, \qquad (4)$$

where $H_{dd}^{sec}$ is the secular part of the Hamiltonian $H_{dd}$.

Conveniently, quantum algorithms start with a pure ground state, where the populations of all states except the ground state are equal to zero. The realization of a pure state in a real quantum system, such as a spin system, requires extremely low temperatures and very high magnetic fields. To overcome this problem, the so-called "pseudopure" state was introduced [25, 26]. The density matrix of the spin system in this state can be partitioned into two parts. The first part of the matrix is a scaled unit matrix, and the second part corresponds to a pure state. The scaled unit matrix does not contribute to observables and is not changed by unitary evolution transformations. Therefore, the behavior of a system in the pseudopure state is exactly the same as it would be in the pure state. We will thus analyze the evolution of the spin system with two groups of initial pseudopure states:

(i) The spin system is initially in a state with all spins up, described by the density matrix:

$$\rho_+(0) = |1\rangle_1 \otimes |1\rangle_2 \otimes ... \otimes |1\rangle_N ; \qquad (5)$$

(ii) The spin system is initially in a state with all spins up except the first spin, which is down; then, the system is described by the density matrix:

$$\rho_-(0) = |0\rangle_1 \otimes |1\rangle_2 \otimes ... \otimes |1\rangle_N , \qquad (6)$$

where $|0\rangle$ represents a spin that is down and $|1\rangle$ represents a spin that is up. These states have no entanglement, since the initial density matrices (5) and (6) are presented as products of the individual spin states. For clusters of dipolar coupled spins, the method of creating highly polarized spin states (5) is based on filtering multiple-quantum coherence of the highest order, followed by a time-reversal irradiation and partial saturation periods [27, 28]. As was demonstrated experimentally [29], the initial state (6) with all spins up except the first spin, which is down, can be prepared using partial saturation and applying a selective Gaussian pulse. With the aim to obtain the analytical solution, we first consider a simple two-spin system, $N = 2$, with equal is the energy difference between the excited and ground states $\omega_{01} = \omega_{02}$ and with equal amplitudes of the irradiation field $\omega_{1,1} = \omega_{1,2} = \omega_1$. In this case, the Hamiltonian $H_{dd}$ describing dipolar interactions in a high magnetic field $H_0$ can be restricted by its secular part:

$$H_{dd}^{\sec} = \sum_{j<k} D_{jk} \left[ I_j^z I_k^z - \frac{1}{4}\left(I_j^+ I_k^+ + I_j^- I_k^-\right) \right], \tag{7}$$

where $D_{jk}$ is the coupling constant between spins $j$ and $k$ (here we used $\hbar = 1$), and $I_j^+$ and $I_j^-$ are the raising and lowering spin angular momentum operators of the spin $j$. The Liouville–von Neumann equation for the density matrix in the rotating frame $\rho_\pm(t)$ has an exact solution. Using the definition of an individual polarization $P_{\alpha,k,\pm} = \text{Tr}\left[\rho_\pm(t) I_k^\alpha\right]$, where $\alpha = x, y, z$, and $k = 1, 2$, we obtain the time dependence of the different components of the polarization:

$$P_{x1\pm} = \mp \frac{2\frac{\omega_1}{D_{12}}}{3\left[1 + \left(\frac{8\omega_1}{3D_{12}}\right)^2\right]} \left( \cos\frac{3D_{12} t \sqrt{1 + \left(\frac{8\omega_1}{3D_{12}}\right)^2}}{4} - 1 \right), \tag{8a}$$

$$P_{y1\pm} = \mp \frac{2\frac{\omega_1}{D_{12}}}{3\sqrt{1+\left(\frac{8\omega_1}{3D_{12}}\right)^2}} \cos\frac{3D_{12}t}{8} \sin\frac{3D_{12}t\sqrt{1+\left(\frac{8\omega_1}{3D_{12}}\right)^2}}{8}, \qquad (8b)$$

$$P_{z1\pm} = \pm \frac{1}{2}\left[\cos\frac{3D_{12}t}{8}\cos\frac{3D_{12}t\sqrt{1+\left(\frac{8\omega_1}{3D_{12}}\right)^2}}{8} \pm \frac{\sin\frac{3D_{12}t}{8}\sin\frac{3D_{12}t\sqrt{1+\left(\frac{8\omega_1}{3D_{12}}\right)^2}}{8}}{\sqrt{1+\left(\frac{8\omega_1}{3D_{12}}\right)^2}}\right], \qquad (8c)$$

where $D_{12}$ is the coupling strength of the dipole-dipole interaction between the nearest spins and "+" and "-" in the two spin system relate to initial ferromagnetic (5) and antiferromagnetic (6) conditions, respectively.

The concurrence of a two-spin system is defined as

$$C = \max\left\{0, 2\lambda - \sum_{k=1}^{4}\lambda_k\right\}, \qquad (9)$$

where $\lambda = \max\{\lambda_1, \lambda_2, \lambda_3, \lambda_4\}$, and $\lambda_1, \lambda_2, \lambda_3$, and $\lambda_4$ are the square roots of the eigenvalues of the operator $R(t) = \rho(t)(\sigma_1^y \otimes \sigma_2^y)\rho^*(t)(\sigma_1^y \otimes \sigma_2^y)$ and $\sigma_k^y$ is the Pauli matrix.

Using the obtained solution for the density matrix, we also found the exact expression for the concurrence as a function of time. For both initial conditions (5) and (6), we found that concurrence and polarization are interlinked by:

$$C_\pm(t) = \sqrt{1 - [2P_{1\pm}(t)]^2}, \qquad (10)$$

where

$$P_{1\pm}(t) = \sqrt{P_{1x\pm}^2 + P_{1y\pm}^2 + P_{1z\pm}^2} \qquad (11)$$

is the polarization of the first spin.

The von Neumann entropy of entanglement is defined as:

$$S = -\text{Tr}(\rho_r \log_2(\rho_r)) \qquad (12)$$

where $\rho_r$ is the reduced density matrix [3, 4].

This entropy is related to the concurrence $C$ by the following equation [30]:

$$S(x) = -x\log_2 x - (1-x)\log_2(1-x) \tag{13}$$

where $x = \frac{1}{2}\left(1 + \sqrt{1-C^2}\right)$.

Using Eqs. (10) and (13), we obtain the relation between the entropy and the individual polarization in the form:

$$S_\pm(t) = 1 - \frac{1+2P_{1\pm}(t)}{2}\log_2(1+2P_{1\pm}(t)) - \frac{1-2P_{1\pm}(t)}{2}\log_2(1-2P_{1\pm}(t)) \tag{14}$$

Expressions (10) and (14) constitute the basis for the experimental investigation of the creation and evolution of entanglement through measurements of individual spin polarization. It is surprising that expression (14) for the von Neumann entropy of entanglement in the ferromagnetic state of dipolar coupling spins exactly coincides with the entropy for non-interacting spins (2).

Fig. 1 illustrates the evolution of the polarization, concurrence and entropy in a two-spin system and shows that the results obtained with the use of individual polarization (using expressions (10) and (14)) and those calculated by using the density matrix coincide at every instant of the time. One method for measuring the individual polarization is described by Lee et al [28]. In the ferromagnetic initial state (5) the polarizations of both spins are the same, and the concurrence and entropy can be expressed through the total polarization of the chain $P_+(t) = P_{1+} + P_{2+} = 2P_{1\pm}$. For antiferromagnetis initial state (6) only individual polarizations of the first and/or second spins relate to the concurrence according expression (10) (Figs. 2a and 2b), while dependence of the concurrence on the total polarization is nonlinear (Fig. 2c).

For initial state (6), the transverse component of dipole-dipole interactions causes mutual flips of the spins and generates entanglement, even without a transverse resonance field, i.e., at $\omega_1 = 0$ (Fig. 3a). Thus, entanglement can appear without an external influence, being due only to internal interactions between spins. It follows from Eqs. (9) that the polarization and, hence, concurrence depend on the amplitude of the irradiation field, $\omega_1$. Using this fact, we can adjust the magnitude of the polarization and control the degree of the entanglement (Fig. 3b).

Generation and control of entanglement for two remote spins in many spin chains ($N > 2$) is the basis for the realization of quantum communication and transfer of information. Entanglement between remote spins can be generated by using selective irradiation [22]. However, despite the selectivity of excitation, due to the transverse component of dipole-dipole interactions, entanglement is shared between all spin pairs in the chain, and the relationship (10) is broken. Only in the weak coupling limit [31], when the difference in precession frequencies $\omega_{0k}$ of interacting spins exceeds the spin-spin interaction strength, can entanglement be localized on a selected spin pair, while the rest of the spins remain separable. In this case, the dipole – dipole Hamiltonian is truncated only to ZZ-terms:

$$H_{dd}^{\sec} = \sum_{j<k}^{N} D_{jk} I_j^z I_k^z. \qquad (15)$$

Spin systems described by Hamiltonians similar to (15) can be found in liquid-state NMR [29]. The numerical simulation of entanglement dynamics in an eight-spin chain with a selectively irradiated pair of spins has shown the validity of the time dependent expression (10), regardless of the location of the spins in the chain. As an example, we give in Fig. 4 the evolution of entanglement between the ends of an eight-spin chain.

We have thus established important relationships between entanglement measures and the order parameter (spin polarization) in spin systems controlled by the NMR technique. With this technique, the dynamic control and measurement of the individual polarization can be realized. This control can be achieved by variation of both the duration and the strength of the irradiated field. We have also shown that it is possible to prepare entangled states between any remote spins in a spin chain, a finding that is of prime importance for the realization of quantum computation and quantum communication.

Our results thus demonstrate that spin systems are an effective tool for studying entanglement dynamics.

**Figures**

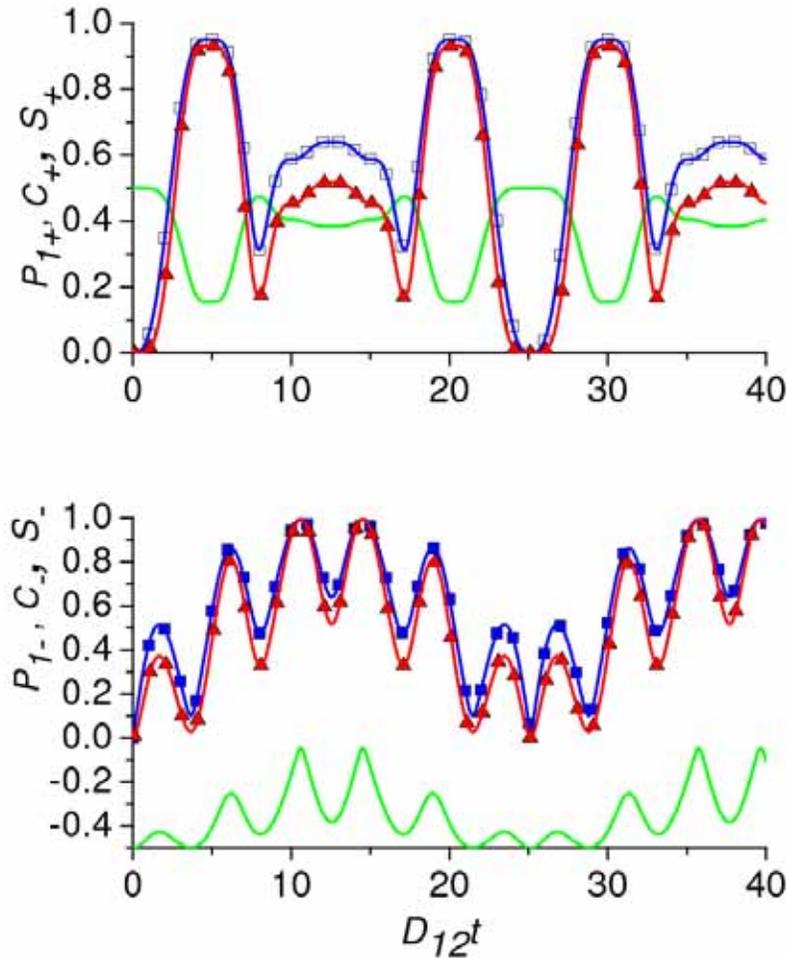

**Figure 1** Time dependences of the individual spin polarization $P_1$ concurrence $C$ and entropy $S$ for different initial states of the two-spin-system at $\dfrac{\omega_1}{D_{12}}=0.5$. (**a**) Initial state (5). (**b**) Initial state (6). Green curves present the polarizations $P_{1+}$ and $P_{1-}$ calculated by using analytical expressions (8) and (11). Blue curves show the concurrences $C_{+}$ and $C_{-}$ calculated by using expression (9), and blue filled squares – for those calculated by using (10). Red lines present the entropies $S_{+}$ and $S_{-}$ calculated by using definition (12) and red filled triangles – calculated by using Eq. (14). The indexes "+" and "-" relate to the initial states (5) and (6), respectively.

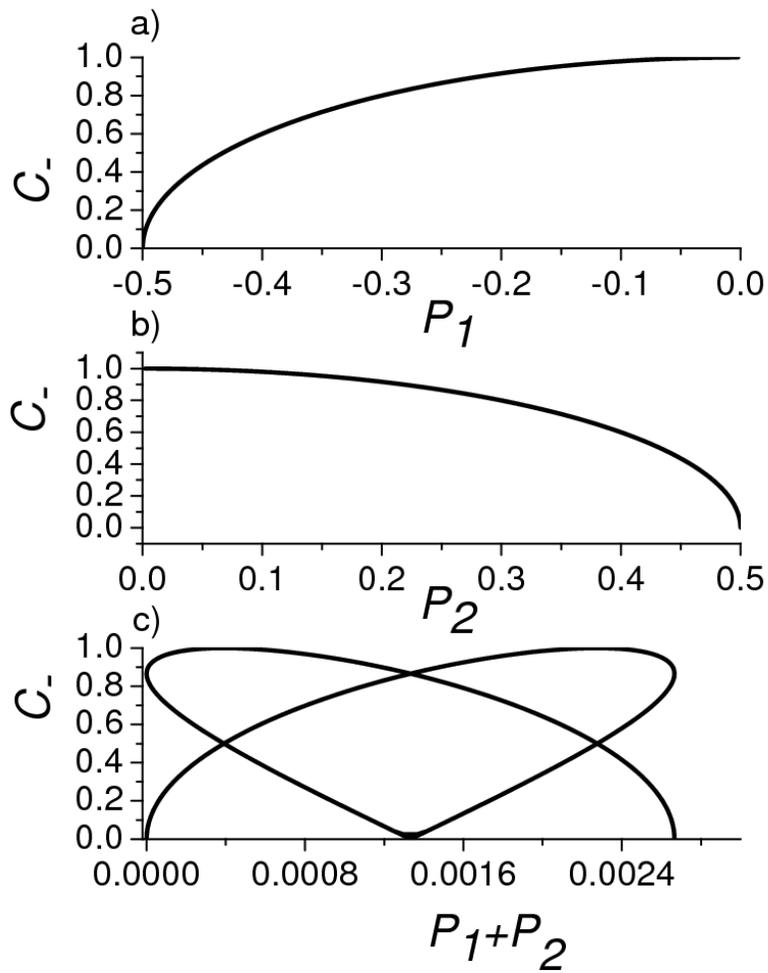

**Figure 2** Concurrence as a function of the polarization for antiferromagnetic initial state (6): (**a**) polarization of the first spin, $P_1$; (**b**) polarization of the second spin, $P_2$; (**c**) the total spin polarization, $P_1 + P_2$.

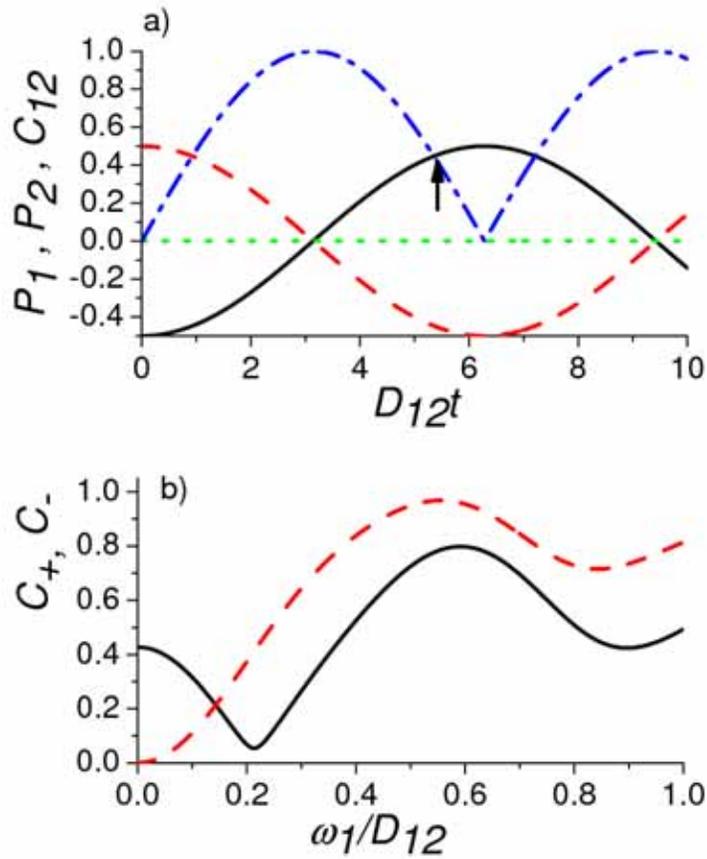

**Figure 3** (a) Concurrence as a function of the irradiation field strength $\omega_1/D_{12}$ at time $D_{12}t = 5.4$ for various initial conditions. Black solid curve: $C_-$ for the initial condition (4). Red dashed curve: $C_+$ for the initial condition (5).

(b) Dynamics of the polarization and concurrence without an irradiation field (at $\omega_1 = 0$) for the initial condition (4). Black dashed curve: polarization $P_{1-}(t)$ of the first spin. Red dashed curve: polarization $P_{2-}(t)$ of the second spin. Green dotted curve: total polarization $P$. Blue solid curve: concurrence $C_-(t)$. The arrow marks the point $D_{12}t = 5.4$

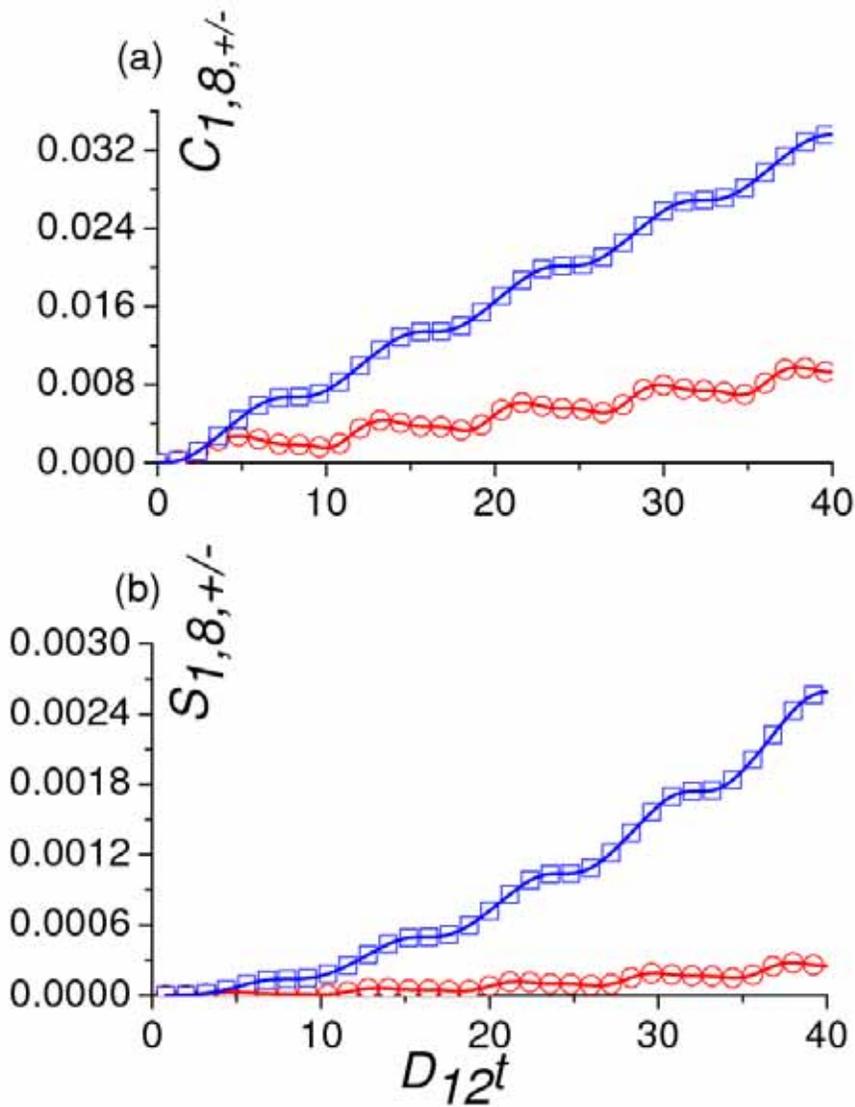

**Figure 4** Generation of entanglement between the ends of an eight-spin chain. (**a**) End-to-end concurrence $C(t)$. (**b**) End-to-end entropy $S(t)$. In both figures, the indexes "+" and "-" relate to the initial states (5) and (6), respectively. Numerical simulations were performed using the MATLAB package: solid curves for the concurrence – calculated using (9), symbols –calculated using (10); solid curves for the entropy- calculated using (12), symbols –calculated using (14).